\documentclass{article}
\usepackage[utf8]{inputenc}
\usepackage{graphicx}
\usepackage{authblk}
\title{A Deep Learning Approach Using Masked Image Modeling for Reconstruction of Undersampled K-spaces
}
\author[1, 2]{Kyler Larsen}
\author[2]{Arghya Pal}
\author[2]{Yogesh Rathi}
\affil[1]{A$\&$M Consolidated High School, College Station, TX, USA}
\affil[2]{Department of Radiology, Brigham and Women’s Hospital, Harvard Medical School, Boston, MA, USA}
\date{August 2022}

\begin{document}

\maketitle

\section{Introduction}
Increasing the efficiency of a magnetic resonance imaging (MRI) scanner can save time, money, and most importantly lives (Jahnke 2019). Most modern day MRIs are effective but have the same major drawbacks; they are expensive and time consuming (Ghadimi et al 2022). The cost can largely affect overall access to MRIs, while the time consumption has a multitude of major effects. Additionally, for patients with pre-existing conditions such as claustrophobia, the time spent in the machine could have negative mental and physical effects, as well as reducing image quality due to the likelihood the patient would move during screening. Patients with certain contraindications such as pacemakers, artificial limbs or hearing aids can only enter MRI machines for a short period of time, meaning the accuracy of their images would be significantly lower than necessary (Ghadimi et al 2022). With the technology of the given age, MRIs are the next target for improvement. 

Recent advances in machine learning (ML) have sparked the development of artificial intelligence (AI) in medicine. The expanding field of disease detection gave way to many studies focusing on observable characteristics in imaging outputs like x-rays or MRIs. For example, a study in 2020 used a VGG-16 transfer learning model to evaluate its performance for detecting COVID induced pneumonia (M. D. Hasan et al 2021). Then came the development of the modern transformer architecture, such as Google’s BERT model. These models were revolutionary in the sense that they learned through context rather than patterns like previous models (Khan et al 2022). Though these models were revolutionary, their advancement also created an innovation break in a rising technique called masked image modeling (MIM). This occurred because most of the powerful transformers like the aforementioned BERT focused on natural language processing (NLP), which created many breakthroughs with regard to language modeling through the use of masked signal modeling (Y. et al 2021). However, due to the severe differences in image and language modeling, many experts focused on the innovation in NLP rather than MIM (Xie et al 2021). It was not until recently that researchers found the first significant breakthroughs in regards to masked image modeling. 

The rationale behind this study develops from the base of how MRI machines work. When a patient enters an MRI machine, a powerful magnetic field is created running parallel to the person’s body. This stimulates the protons in the patient’s body, causing them to move into the direction of the magnetic field (Berger 2002). Then, a radiofrequency current is pulsed through the patient’s body, causing the protons to pull against the magnetic field. When the radiofrequency is turned off, the MRI machine is able to detect the protons returning to their equilibrium state (Berger 2002). The machine uses a series of sensors arranged in coils, with each sensor focusing on a different part of the body. The scan obtained from each individual sensor, or slice, is then combined with all the other slices from that coil. After obtaining each image’s full coil scan, the machine then uses the root-sum-squares method to combine the coils into the complete image (Hansen et al 2016) This process takes time, and if the patient moves at all, the scan could become inaccurate. If instead only a percentage of these sensors were active, the time per scan could decrease by as much as 65$\%$ (Gassenmaier et al 2021) This produces a raw data form called a k-space, which represents the spatial distribution of information acquired during the scanning process (Moratal et al 2008). 
$$RMSE = \sqrt{\sum_{i = 1}^{n}\frac{(\hat{y}_i - y_i)^2}{n}}$$

The code used behind this study is a modified version of the Simple Masked Image Modeling (SimMIM) framework found in the paper cited to Xie et al 2021\footnote{https://github.com/microsoft/SimMIM}. The SimMIM code illustrated the first simple framework (shown in Figure 1) for the technique of masked image modeling. This technique of masked image modeling has potential to make the next breakthrough in the field of disease detection, meaning it is essential to understand and develop this code (Chen et al 2022). The SimMIM is revolutionary in comparison to other baseline MIM models since it truly finds the simplest and most efficient model feasible; it utilizes raw pixel regression, extremely light (linear) prediction heads, and random masking (Xie et al 2021). Previous studies using machine learning to improve MRIs have focused largely on reconstructing the full MRI image using older architectures such as UNet (Ernst et al 2021, Ronneberger et al 2015). These models are effective, but to increase possibilities for improvement, newer and more advanced architectures must be evaluated (Zeng et al 2021). \\ \\
\begin{figure}
    \centering
    \includegraphics[width=320pt]{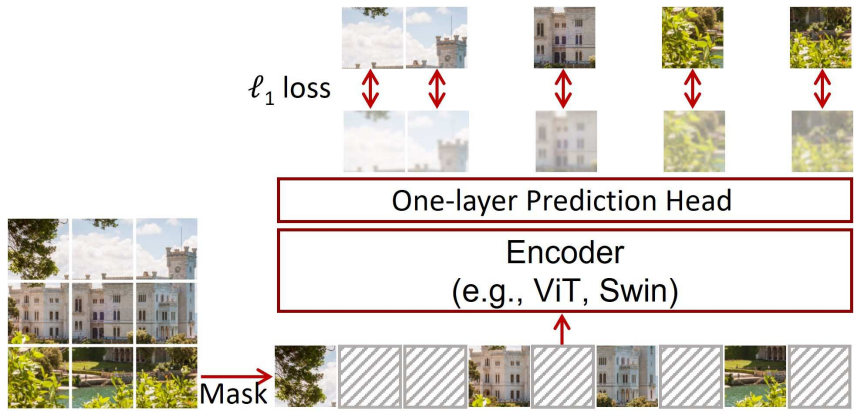}
    \caption{Simmim Architecture Diagram - The masked images are passed through the encoder, then to the decoder before being evaluated on L1 loss and then classification (Xie et al 2019)}
\end{figure}
The purpose of this study was to develop and evaluate a model that could receive input in the form of an undersampled k-space and produce an accurate MRI image. The goal of this study was to develop such a model that could perform with fully sampled k-space level accuracy while utilizing an undersampled space. The significance of this study is that it would be the first ever model to utilize a MIM-based model to streamline the process of MRI scanning. It would provide a baseline model and valuable information regarding the feasibility of using MIM in the medical field.\\

\section{Methods}
\subsection{Dataset}
This study makes use of the open source fastmri dataset (Zbontar et al 2019). Specifically, this study uses only knee MR images and k-spaces from the validation set of images. The dataset includes Contains fully sampled knee MRIs from 1500 patients obtained from 1.5 and 3.0T scanners. The dataset also contains more than 10,000 DICOM images with coronal and axial proton density with and without fat suppression (Figure 2), T2 weighted with fat suppression and sagittal proton density. The dataset was downloaded as multiple h5 files, meaning a custom dataloader had to be built to pass the data into the model. The model was unable to process h5 datasets or images, meaning the images were saved and passed through as PNG images. These images were not split into seperate categoires during training or validation. Only slight augmentations such as cropping for resizing were applied to the data for the purpose of preserving image details. The data was split 80/20 for train and test.

\begin{figure}[!ht]
    \centering
    \includegraphics[width=320pt]{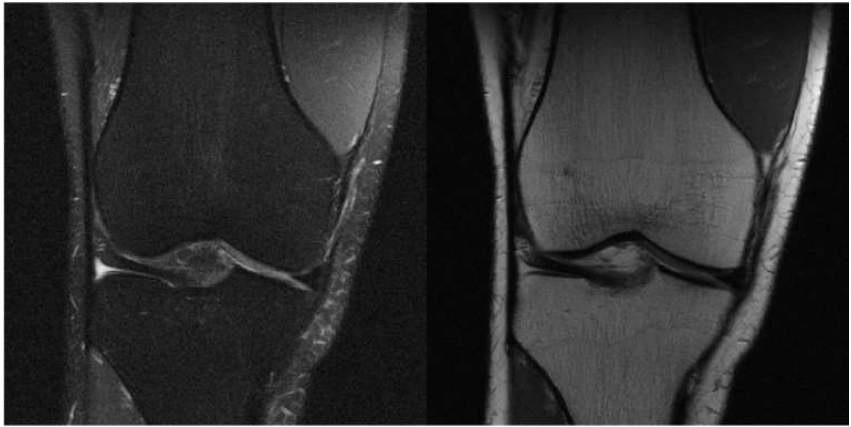}
    \caption{Knee images from fastmri dataset, left is proton density image with fat suppression, right is proton density image without fat suppression (Zbontar et al 2019)}
\end{figure}

\subsection{Architecture Overview}
The model used in this study is a modified version of the SimMIM model. The model features two primary encoders for classification; the vanilla Vision Transformer (ViT), and then Shifted-Window transformer (Swin). The SimMIM utilizes an encoder-decoder architecture, with the encoder being either of the aforementioned models, and the decoder being a sequential model with basic 2D convolutional layers. \\
\begin{figure}[!ht]
    \centering
    \includegraphics[width=320pt]{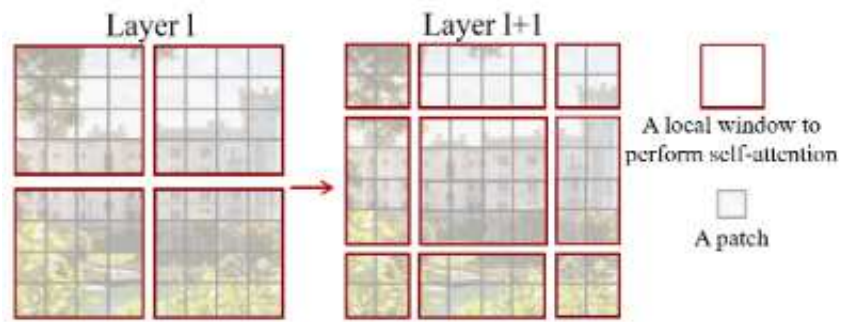}
    \caption{Shifted Window pattern, shown for progressing layers (Liu et al 2019)}
\end{figure}

The Swin Transformer is a type of vision transformer that classifies based on a shifted window algorithm. The Swin creates a hierarchical representation by starting from small sized patches and gradually merging neighboring patches. As the layers progress (Figure 3), the window for self-attention calculation (red) "shifts" , resulting in new windows. These new windows cross the boundaries of the old windows, forming attention between them (Liu et al 2021). This method works well for classification, but is not optimized for prediction. The baseline model performed reasonably well for reconstruction (See Figure 11), but changes were made to improve prediction performance.
\begin{figure}[!ht]
    \centering
    \includegraphics[width=320pt]{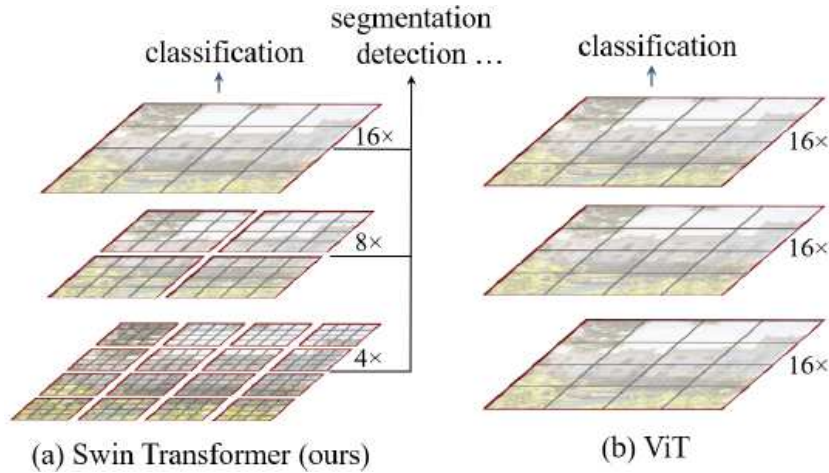}
    \caption{Comparison of architectures for Swin vs ViT (Xie et al 2021)}
\end{figure}

The ViT transformer works similarly to the Swin transformer. It receives input in the form of a 1D sequence of token embeddings (Dosovitskiy et al 2021). The same methodology of patch embeddings as the Swin transformer are applied (Figure 4). Position embeddings are added to the patch embeddings to retain positional information. The difference between this model and the Swin is that the window for self-attention caluclation does not shift between layers (Liu et al 2021). While this approach suffices for classification, it becomes a drawback for image reconstruction since the details of the image are not preserved during extraction.

\subsection{Experimentation}
For experimentation in this study, the primary changes were made with the hyperparameters in the encoders. To begin, the model first had to be transferred from classification optimized to prediction optimized. The number of classes was reduced to 1, batch size reduced to 1, and the several patch sizes and encoder strides were experimented with. The model found decent success with a patch size of 1 for original embeddings, and 1 for model initialization, a significant decrease from the original values of 4 for each. The encoder stride was also changed, and the model found the most success with an encoder stride of 32 (default) since the model had already been optimized with that value. The number of epochs was held at a steady 100 for all trials. \\
Before the data was passed into the model, it was resized to fit the model. Originally, the data had the shape of 384 $\times$ 680, but the model only uses square images. To control for this, this study only used center crop to resize the image to 192 $\times$ 192. Augmentations such as random flip or recolor were removed for training, since they are for classification purposes and do not preserve any of the necessary details. For normalization, trials were first run with default ImageNet normalization values (model default), then run with image specific mean and standard deviation values, before being removed completely. Ultimately, we decided to remove normalization from the augmentations since the dataset was relatively small and normalization removes some of the fine details. After the augmentations were made, the images were passed through a masking function. This function was used to simulate under sampling of the k-space, since the removal of input channels would train the model in similar methods. \\
It is important to note that for this study, the saved images were in the format of 'RGBA' instead of the typical 'RGB'. For the models, which use 'RGB' base, many of the functions (especially weights) were optimized and even hard coded for 3 input channels only. This meant we had to create a custom dataloader and modify each function, making sure to scale each augmentation, as well as rewriting the weights and gradients to accommodate the batch size and input channel change. The model was evaluated on the metrics of loss, gradient normalization, and structural similarity. Specifically, this study used L1-loss to determine base similarities between the reconstructed images. Gradient normalization was used to indicate the training performance of the model, making sure that during pretrain the model reached optimal state. 
$$SSIM(x,y) = \frac{(2\mu_{x}\mu_{y} + c_1)(2\sigma_{xy} + c_2)}{(\mu_{x}^{2} + \mu_{y}^{2} + c_1)(\sigma_{x}^{2} + \sigma_{y}^{2} + c_2)}$$
Structural similarity (SSIM) was used as the primary statistic for evaluation since it demonstrates a direct comparison between the image and the reconstructed image. Because it's primary non machine learning use is to determine drops in image quality (as well as image similarity), it functioned for both image quality and similarity measurement (Blaimer et al 2004). The primary drawback of skimage's SSIM function is that it requires the images to be passed in as grayscale; however, in this case, our images are already grayscale, meaning minimal detail was lost.

\section{Results}
The model was first evaluated on combined image and metric results, then evaluated only on the obtained metrics when the model reached the fine-tuning stage.
\subsection{Image Results}
For the Swin transformer, the model saved reconstructed images with structural similarity values over 0.995, and for the ViT transformer, the model saved images with values over 0.60. Originally, the model was also going to evaluated based on low validation loss values ($<$0.001), but by image and SSIM comparison, we concluded that low validation loss was not a proper measurement for image similarity. Based on images similar to the ones shown in Figure 5, it was determined that the patch size of the model and the encoder stride were too large for detail reconstruction (as evidenced by the appearance of solid color patches). 
\begin{figure}[!htt]
    \centering
    \includegraphics[width=140pt]{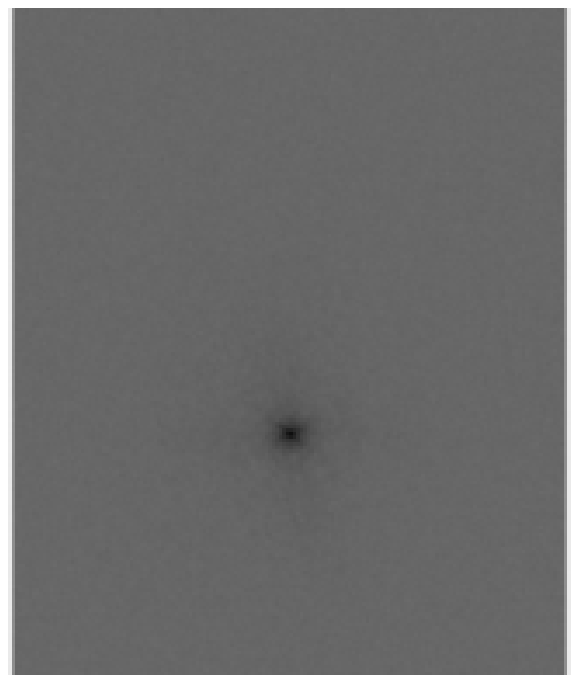}
    \includegraphics[width=142pt]{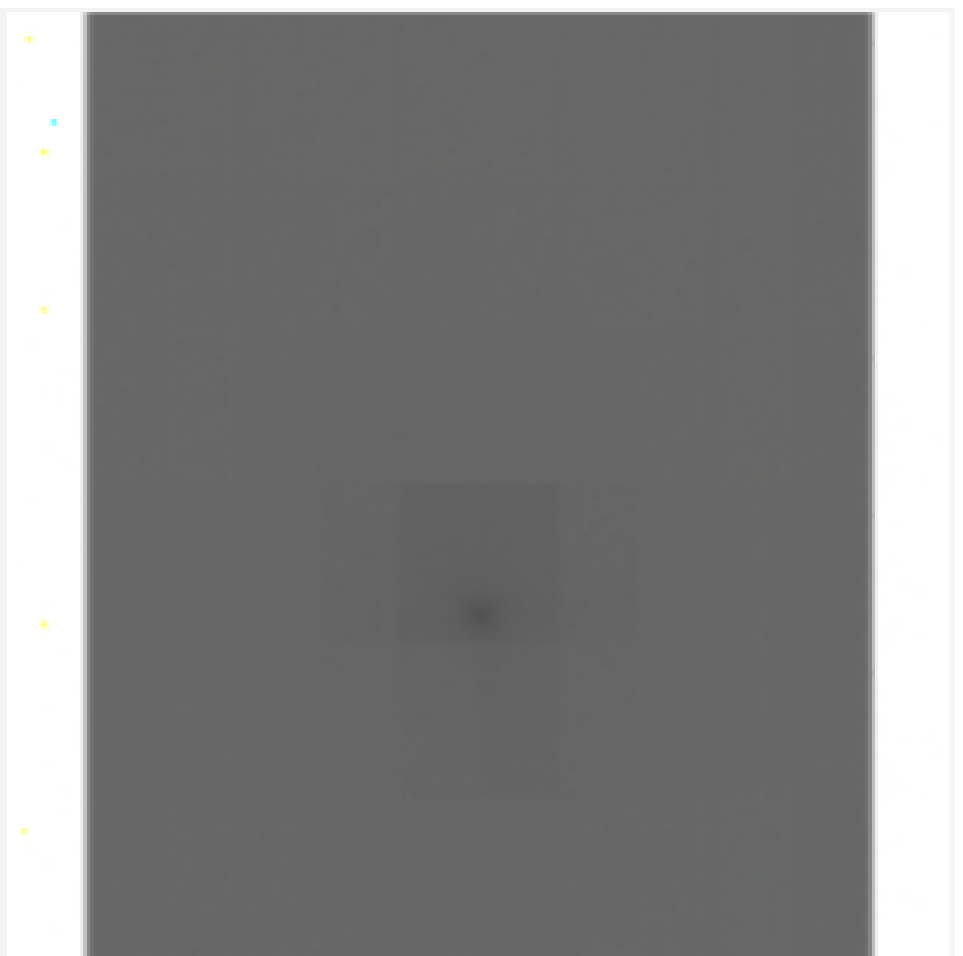}
    \caption{Original fully sampled k-space versus reconstructed k-space passed through with Swin encoder}
\end{figure}

For the above images, the specific SSIM value was 0.9956, showing that even a 99.56$\%$ structural similarity can miss key features. However, the Vision transformer performed significantly worse. 
\begin{figure}[h!]
    \centering
    \includegraphics[width=130pt]{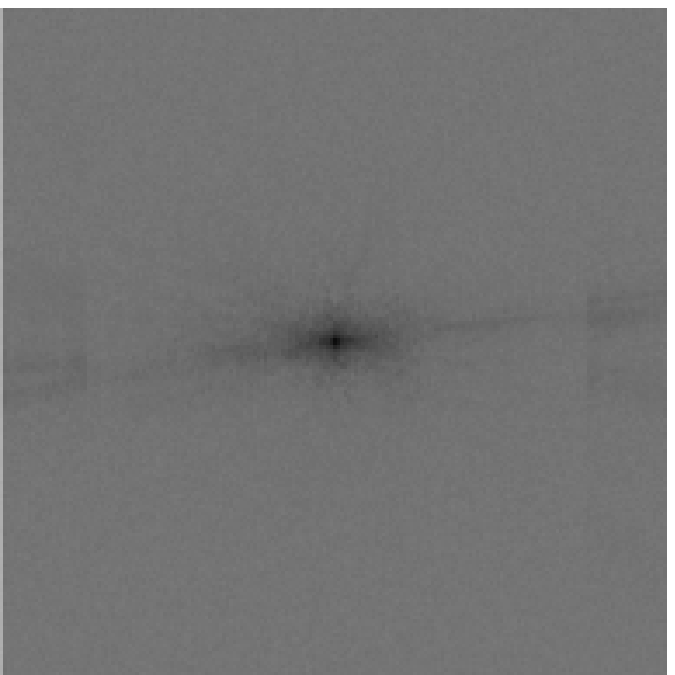}
    \includegraphics[width=130pt]{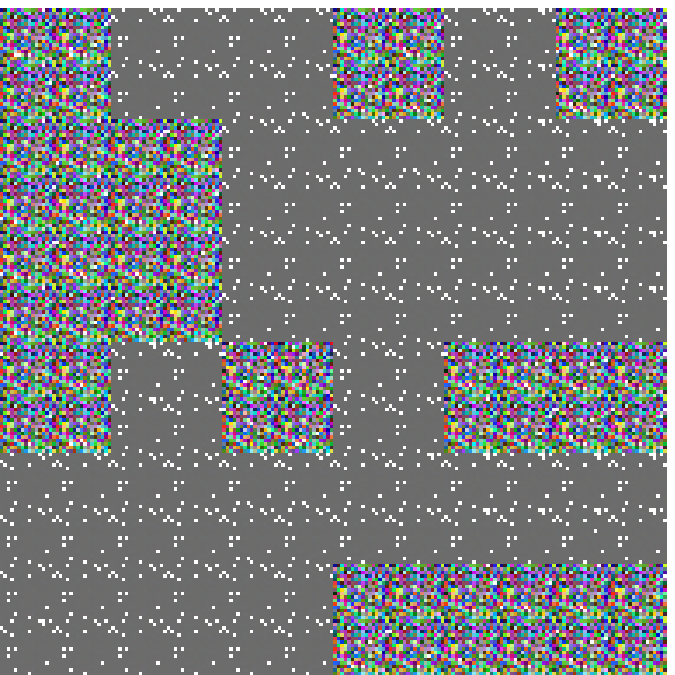}
    \caption{Original fully sampled k-space and reconstructed k-space passed through with ViT encoder}
\end{figure}

It was to be expected that the Swin would outperform the ViT transformer in reconstruction (Figure 6); however, it was still interesting to see by how much it under performed in comparison to the Swin. 

\subsection{Graphical Results}
Before the model reached the fine-tuning stage, metrics such as training and gradient norm were collected and graphed to ensure the model never overfit and reached a steady state before being saved for fine tuning. 
\begin{figure}[!ht]
    \centering
    \includegraphics[width=350pt]{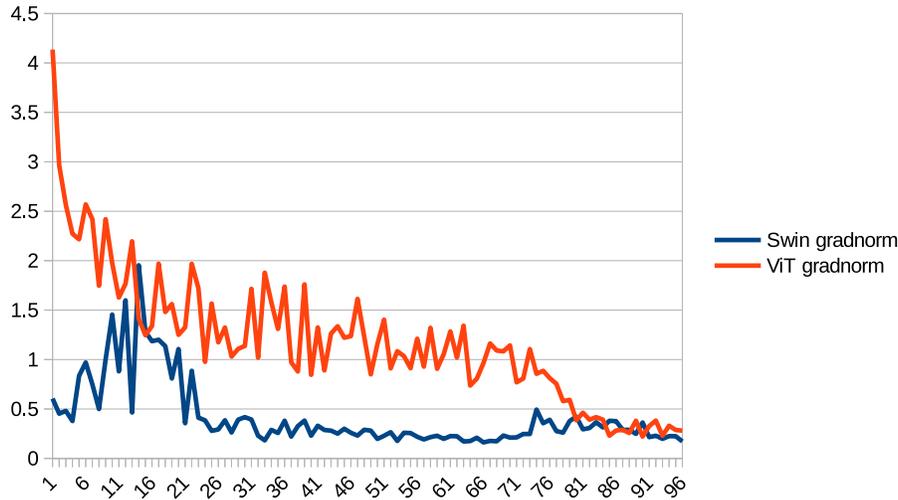}
    \caption{Gradient norm for each encoder model, shown for first 97 epochs}
\end{figure}
From the gradient normalization data shown in Figure 7, it was determined the models reached a steady state while completing the training cycle. Investigations into overfitting were not undertaken until the fine-tuning portion of the model. For this, a graph of validation and training loss was constructed during the fine-tuning cycles. These graphs were used to determine if the model was overfitting by examining the trend of the validation loss.
\begin{figure}[!ht]
    \centering
    \includegraphics[width=350pt]{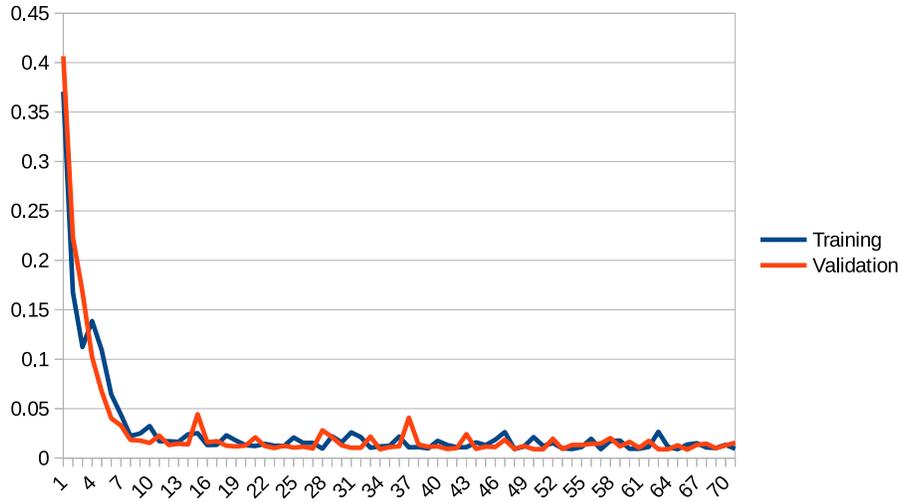}
    \caption{Validation and Training loss for Swin encoder-based SimMIM}
\end{figure}

Figure 8 illustrates the the validation and training loss of the Swin encoder-based SimMIM. Since the validation loss continued to change with the training loss, it can be ascertained that the model was not memorizing the data, but rather using the features extracted by the training during each new validation cycle. Figure 9 (below) illustrates the validation and training loss of the ViT encoder-based SimMIM. Unlike the Swin base, the validation accuracy does not significantly shift with the training loss. This shows that since the validation loss is not responsive to the training, it is instead just memorizing the data without improving. 
\begin{figure}[!ht]
    \centering
    \includegraphics[width=350pt]{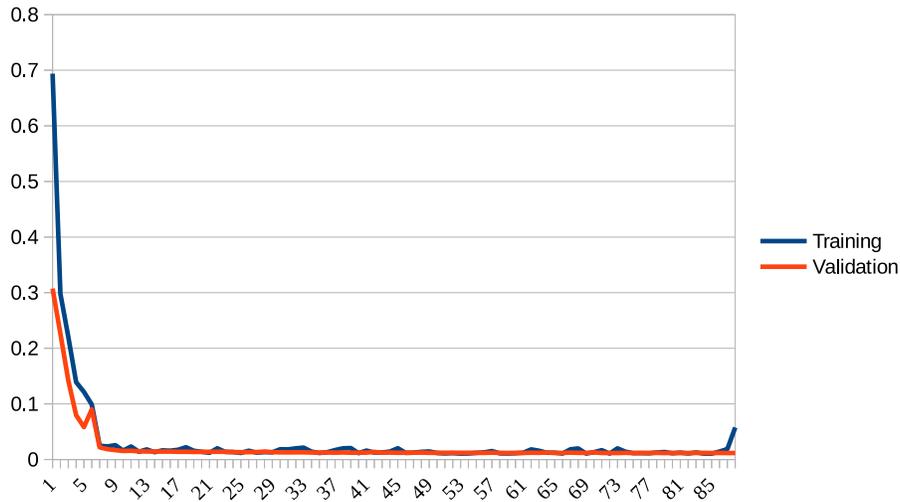}
    \caption{Validation and Training loss for ViT encoder-based SimMIM (first 70 epochs), the flat-lining validation loss illustrates a non-responsiveness to the training loss, meaning the model is memorizing the data as opposed to learning.}
\end{figure}
For the SSIM portion of data collection, graphs were made between both validation and loss, as well as between both of the encoders. It is important to note that while SSIM seemed to have an inverse correlation with L1-loss, their relationship is less substantial than expected. While both the Swin and ViT encoder models approached validation loss values of $<$0.01, only the Swin encoder exceeded SSIM values of 99$\%$, while the ViT encoder hovered around 60$\%$.
\begin{figure}[!ht]
    \centering
    \includegraphics[width=340pt]{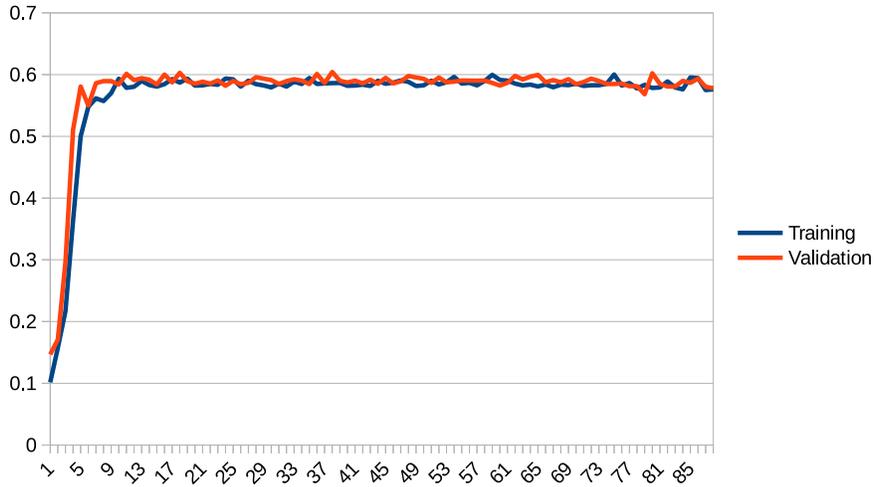}
    \caption{ViT training vs validation SSIM.}
\end{figure}
\begin{figure}[!ht]
    \centering
    \includegraphics[width=340pt]{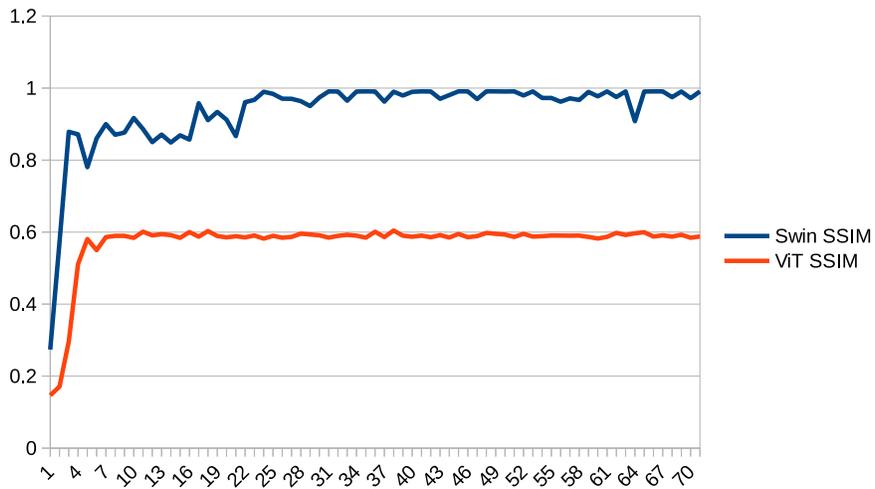}
    \caption{ViT vs Swin encoder based SSIM}
\end{figure}

Figure 11 illustrates the ViT encoder's maximum SSIM values. While these values hovered consistently around 0.60, Figure 10 illustrates the validation loss values to be $<$0.05 during the same epochs. In comparison, the Swin encoder has similar loss values (though fluctuating more) with significantly higher sturctural similarity. The Swin transformer outperformed the ViT by SSIM values of over 49$\%$. This structural similarity value illustrates a reasonable success in reconstruction, although some of the fine details were lost. 
\section{Discussion}
The model performed well by the standard of structural similarity to the original fully sampled k-space. SSIM scores consistently over 99.5$\%$ and k-space images that reconstruct the extremeties well illustrate the model completed the reconstruction task.
It is worth noting the lack of reconstructed MR image in this study, only the reconstructed k-space. This happens because during model processing, the complex dimension of the PNG images was reset, meaning an inverse Fourier Transform consistently yielded incorrect results or dimension errors. Because this study focused on reconstructing the k-space as opposed to the full MRI, this means a full a full, reconstructed MR image was unattainable on the current experimental design. 
\subsection{SwinRecNet}
Attempts were made to incorporate other, more reconstruction focused models as encoders for this study. One such model was the SwinRecNet developed in the study by Pan et al 2021 (Figure 12). 
\begin{figure}[!ht]
    \centering
    \includegraphics[width=340pt]{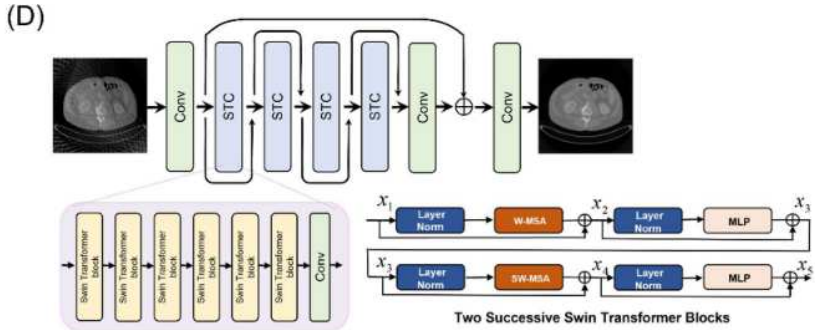}
    \caption{SwinRec Architecture Diagram (Pan et al 2021)}
\end{figure}

The SwinRec functions very similarly to the SwinTransformer, but it optimizes reconstruction instead of classification. It does this by using an edge enhancement reconstruction sub-network for recovering the initially-reconstructed MR image. The problem with using this architecture is that it is also optimized for 'RGB' images, and was developed for a different model MistNet. This means transferring the model to to be compatible with the SimMIM made it lose certain features, or compatibility as a whole. Ultimately the model was not included in this study due to the loss of performance it found when being made compatible. \\
\subsection{Augmentation}
This study also examined the effects of augmentation on reconstruction. Data augmentation such as recoloring and interpolation were removed for detail preservation reasons. These augmentations cause significant changes to the images, becoming counterproductive in balance with overfitting reduction.
However, slighter augmentations such as normalization, randomflip, and randomcrop were tested to see if the training metrics would improve as a result. Data was gathered for both augmented data models and models where the data faced no augmentation except center crop (for size constraints).
\begin{figure}[!ht]
    \centering
    \includegraphics[width=340pt]{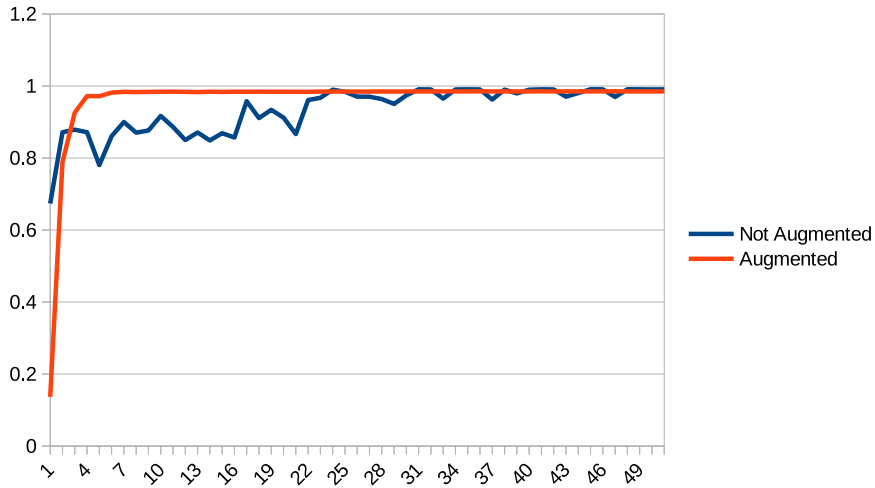}
    \caption{Augmentation Data for Swin Transformer (first 50 epochs)}
\end{figure}

Figure 13 illustrates the relationship between the SSIM of model with augmented data and the SSIM of a model without. Interestingly, the model's SSIM appears to increase much quicker with augmentation, but improves less over time than the model with non-augmented data. Since augmentation is used typically to reduce  overfitting chances and keep the model learning, this is an interesting result that the augmented data model appears to have a constant SSIM value for epochs 10-50. This finds that for reconstruction, not augmenting the data appears to have a greater effect on training the model. This is inversely true in comparison to many classification networks, although it makes sense because augmentation should not be applied for many basic reconstruction networks due to image alterations.

\subsection{Limitations}
Because of the fine details required between k-space and image conversion, the reconstructed image shown in Figure 5 (right) will likely produce an inaccurate MR image. Additionally, this study only utilized a relatively small dataset of ~7500 images, in comparison to other studies that use datasets with millions of images. Of course, this would probably help the overall reconstruction, since the model would have more images to train on, but it is still important to note that this study was conducted on a small sample set. Lastly, the masking in this study was only simulated through a masking function. In a real world masked k-space, the masking might function differently, meaning the model would need to be tested on actual masked k-spaces to confirm the results for reconstruction.
\section{Conclusion}
The goal of this study was provide information regarding the feasibility of using MIM for MRI reconstruction, and to modify an existing model to optimize this approach for k-space reconstruction to create a fully-sampled k-space from a masked k-space. Changes to the masking and patch embed functions as well as to the various hyperparameters were made to optimize the model for reconstruction rather than classification. The fine-tuned model achieved a structural similarity of over 99.5$\%$ with the Swin encoder, although the reconstructed image lacked certain details. The reconstructed k-space would likely not produce a completely accurate MR image, but the results are promising for using masked image modeling for reconstruction. Further studies could take advantage of reconstruction optimized encoders such as the aforementioned SwinRecNet for further increase to image quality. Additionally, reworking the dataloaders to accept h5 files (so that the model can pass through non PNG images) would allow the actual reconstructed MR image to be accessible. Overall, this study concludes that with some adjustments, masked image modeling can be useful for MR image and k-space reconstruction.
\section{Acknowledgements}
Special thanks to Dr. Kang-Ik Cho for his help in revisions and coding setup, and to the Psychiatry and Neuroimaging Laboratory at Harvard Medical School for use of their GPU and storage, as well as motivations. \\
\\
Link to Github:\\
https://github.com/Aopsmath99/MIMMRI
\newpage
\section{Literature Cited}
Berger, A. (2022). Magnetic Renosance Imaging. National Library of Medicine. https://www.ncbi.nlm.nih.gov/pmc/articles/PMC1121941/.\\ \\
Blaimer, M., Breuer, F., Mueller, M., Heidemann, R. M., Griswold, M. A., \& Jakob, P. M. (2004). SMASH, SENSE, PILS, GRAPPA: how to choose the optimal method. Topics in magnetic resonance imaging : TMRI, 15(4), 223–236. https://doi.org/10.1097/01.rmr.0000136558.09801.dd \\ \\
Chen, Z., Agarwal, D., Aggarwal, K., Safta, W., Balan, M. M., Sethuraman, V., \& Brown, K. (2022). Masked Image Modeling Advances 3D Medical Image Analysis. doi:10.48550/ARXIV.2204.11716\\ \\
Ernst, P., Chatterjee, S., Rose, G., Speck, O., \& Nürnberger, A. (2021). Sinogram upsampling using Primal-Dual UNet for undersampled CT and radial MRI reconstruction. doi:10.48550/ARXIV.2112.13443 \\ \\
Gassenmaier, S.; Küstner, T.; Nickel, D.; Herrmann, J.; Hoffmann, R.; Almansour, H.; Afat, S.; Nikolaou, K.; Othman, A.E. Deep Learning Applications in Magnetic Resonance Imaging: Has the Future Become Present?. Diagnostics 2021, 11, 2181. https://doi.org/10.3390/diagnostics11122181 \\ \\
Ghadimi M, Sapra A (2022). Magnetic Resonance Imaging Contraindications. Treasure Island (FL): StatPearls Publishing. https://www.ncbi.nlm.nih.gov/bo\\oks/NBK551669/ \\ \\
Hansen, M., \& Kellman, P. (2014). Image reconstruction: An overview for clinicians. Journal Of Magnetic Resonance Imaging, 41(3), 573-585. https://doi.org/\\10.1002/jmri.24687\\ \\
Jahnke, A., 2022. Speeding Up MRI Scans to Save Lives. 
The Brink. Available at: $<$https://www.bu.edu/articles/2019/making-mri-scans-faster/$>$ \\ \\
Khan, S., Naseer, M., Hayat, M., Zamir, S. W., Khan, F. S., \& Shah, M. (2022). Transformers in Vision: A Survey. ACM Computing Surveys. doi:10.1145/35052\\44 \\ \\
Liu, Z., Lin, Y., Cao, Y., Hu, H., Wei, Y., Zhang, Z., … Guo, B. (2021). Swin Transformer: Hierarchical Vision Transformer using Shifted Windows. doi:10.48550/ARXIV.2103.14030 \\ \\
M. D. Kamrul Hasan, Sakil Ahmed, Z. M. Ekram Abdullah, Mohammad Monirujjaman Khan, Divya Anand, Aman Singh, Mohammad AlZain, Mehedi Masud (2021). Deep Learning Approaches for Detecting Pneumonia in COVID-19 Patients by Analyzing Chest X-Ray Images", Mathematical Problems in Engineering, vol. 2021, Article ID 9929274, 8 pages. https://doi.org/10.1155/2021/99292\\74 \\ \\
Moratal, D., Vallés-Luch, A., Martí-Bonmatí, L., \& Brummer, M. (2008). k-Space tutorial: an MRI educational tool for a better understanding of k-space. Biomedical imaging and intervention journal, 4(1), e15. https://doi.org/10.2349\\/biij.4.1.e15 \\ \\
Pan, J., Zhang, H., Wu, W., Gao, Z., \& Wu, W. (2021). Multi-domain Integrative Swin Transformer network for Sparse-View Tomographic Reconstruction. doi:10.48550/ARXIV.2111.14831 \\ \\
Ronneberger, O., Fischer, P., \& Brox, T. (2015). U-Net: Convolutional Networks for Biomedical Image Segmentation. Arxiv. doi:10.48550/ARXIV.1505.04597 \\ \\
Xie, Z., Zhang, Z., Cao, Y., Lin, Y., Bao, J., Yao, Z., … Hu, H. (2021). SimMIM: A Simple Framework for Masked Image Modeling. doi:10.48550/ARXIV.2 \\ 111.09886\\ \\
Y. -A. Chung et al. (2021) w2v-BERT: Combining Contrastive Learning and Masked Language Modeling for Self-Supervised Speech Pre-Training, 2021 IEEE Automatic Speech Recognition and Understanding Workshop (ASRU), pp. 244-250, doi: 10.1109/ASRU51503.2021.9688253.\\ \\
Zbontar, J., Knoll, F., Sriram, A., Murrell, T., Huang, Z., Muckley, M. J., … Lui, Y. W. (2018). fastMRI: An Open Dataset and Benchmarks for Accelerated MRI. doi:10.48550/ARXIV.1811.08839 \\ \\
Zeng, G., Guo, Y., Zhan, J. et al. A review on deep learning MRI reconstruction without fully sampled k-space. BMC Med Imaging 21, 195 (2021). https://doi.org/10.1186/s12880-021-00727-9\\ \\

\end{document}